\documentclass[10pt]{article}
\usepackage{amssymb}
\usepackage{amsmath}
\usepackage{amsthm}
\usepackage{latexsym}
\usepackage[dvips]{epsfig}
\usepackage{mathrsfs}
\usepackage{eufrak}

\theoremstyle{plain}

 \font\tenscr=rsfs10 scaled1100
\font\sevenscr=rsfs7 
\font\fivescr=rsfs5 
\skewchar\tenscr='177 \skewchar\sevenscr='177 \skewchar\fivescr='177
\newfam\scrfam
\textfont\scrfam=\tenscr \scriptfont\scrfam=\sevenscr
\scriptscriptfont\scrfam=\fivescr

\newcommand{\co}{\stackrel{+}{\mathtt{a}}}
\newcommand{\ao}{\stackrel{-}{\mathtt{a}}}
\newcommand{\pd}[2]{\frac{\partial #1}{\partial #2}}

\begin{document}


\title{\textbf{Fluctuation-driven heterogeneous chemical processes}}

\author{{\Large Christiane M Losert-Valiente Kroon}\thanks{E-mail address:
{\tt c.kroon@ucl.ac.uk}} \ and {\Large Ian J Ford} \\ Department of
Physics and Astronomy\\
London Centre for Nanotechnology
\\University College London, Gower Street, London WC1E 6BT, UK.
\vspace{5mm} }

\maketitle

\begin{abstract}
We explore a new framework for describing the kinetics of a
heterogeneous chemical reaction where two particles of the same
chemical species form a reaction product of another chemical species
on the surface of a seed particle. Traditional treatments neglect
the effect of statistical fluctuations in populations. We employ
techniques in a manner analogous to the treatment of quantum systems
to develop a stochastic description of processes beyond the mean
field approximation.
\end{abstract}

\section{Introduction}

Traditional models employing the evolution of the mean population of
species in a system provide a good enough description of various
physical processes such as heterogeneous chemical reactions and
heterogeneous nucleation of aerosols as long as the average particle
number is large. However, when the mean populations involved are
small we expect to see a significant deviation from the solution to
the classical equations.
\\ \\
As an example consider the chemical reaction of two hydrogen atoms
on the surface of a dust particle in interstellar space
\cite{Kle;06}. Atoms can be adsorbed onto and evaporated from the
grain particle. The adsorbed molecules will diffuse on the surface
of the seed and eventually collide with another reactant to form
diatomic hydrogen. Under interstellar conditions the rate of the
adsorption of atoms will be small compared to the reaction rate. If
the mean population of hydrogen atoms is small, statistical
fluctuations brought about by gain and loss, and by the random
diffusion of atoms on the surface of the dust particle, are
important. Yet the traditional model does not include the treatment
of such fluctuations. Several attempts have been introduced to
resolve that problem; among them the Monte Carlo approach, the
modified rate approach, the direct master equation approach and the
Gauge Poisson representation approach
---see \cite{Her;01, HerShem;03} and \cite{Dru;04}. Related studies
have been carried out in \cite{DelFraHilKit;02, HocZorMor;06}.
Analytical solutions to a master equation approach have been found
for the steady state case\cite{BihFur;01, Gre;01, LusBhaFor;03}. Our
work is a first step to an alternative approach for describing the
kinetics of a heterogeneous chemical reaction which can be extended
to other areas where a similar problem occurs, for example when
computing the rate of nucleation processes taking place on small
particles. Again, the number of adsorbed molecules is small and
statistical fluctuations need to be taken into account. In
\cite{LusBhaFor;03} the steady state solution was studied employing
a description using master equations instead of mean population
dynamics. A nucleation rate lower than the classical equations
predicted was obtained. Therefore, it is important to develop a
model by which we replace the classical equations with a set of
stochastic equations.
\\ \\
For the correct treatment of population fluctuations we employ
methods of Quantum Field Theory \cite{Doi;76I, Doi;76II}. Starting
from a master equation we introduce a spatial lattice where the
microstates of the system correspond to a set of occupation numbers
at each lattice site. A Fock space is constructed using annihilation
and creation operators at each lattice site. By means of this set-up
it is easy to show that the master equation is equivalent to a
Schr\"odinger equation with imaginary time. This enables us to
employ techniques originally developed in order to describe a
quantum mechanical system where the fluctuations are due to a
quantum uncertainty. We obtain the average particle population of
the classical many-body system by developing a mechanism for
computing expectation values of observables analogous to Feynman's
path integral formulation \cite{KlaSka;85, Pel;85}. In this paper we
will not concern ourselves with renormalisation group analysis
although studies in that direction have been undertaken by various
people, for example \cite{HowCar;95, Lee;94, ReyDro;97}. Introducing
a stochastic variable \cite{ReyCar;99}, a mathematical trick ---for
a nice review paper we refer to \cite{TaeHowVol;05}--- helps with
the evaluation of the expression for the expectation values. The
complex fluctuating solutions to a set of constraint equations,
which are stochastic partial differential equations, are then
averaged over all realisations of the stochastic noise. For
numerical investigations, the solutions to the constraint equations
can be generated by various numerical schemes \cite{KloPla;92}. The
path integral average is computed using Monte Carlo
methods\cite{Met;53}. The Code is written in C and computations do
not take longer than a few seconds up to a few minutes running on a
standard laptop.
\\ \\
In the sequel we will concentrate on a heterogeneous chemical
reaction where two reaction partners of the same sort of particle
type react on the surface of a seed particle to form a reaction
product.
\\ \\
For readers more interested in the physical content than in the
mathematical details of the formalism we recommend to skip the first
few pages and instead have a look at the most important equations on
page 8 and continue from there. Equation \eqref{average} which is a
path integral average (PIA) gives the average particle density of
the various chemical species once the complex, fluctuating solutions
to the constraint equations \eqref{Ll2AAC} and \eqref{LlAAC} have
been found and inserted into the PIA. Together with the correlations
\eqref{corr} for the stochastic noise occuring in the PIA this forms
a complete set of equations.

\section{Mathematical Techniques}
\subsection{Master Equation And Schr\"odinger-like Equation}

We concentrate on chemical reactions of type $A+A \longrightarrow
C$, that is situations in which the atoms adsorb onto grain
particles where they can associate with themselves to produce
diatomic molecules. If the number of reactive species on an
individual grain is small, traditional rate equations will fail to
accurately describe the diffusive chemistry occuring on the surface
of the grain particle. We start our investigations by determining a
master equation that describes such a heterogeneous chemical
process.
\\ \\
A $\emph{general master equation}$ can be written as
\begin{equation}
\frac{dP(m)}{dt}=\sum_n T_{n\rightarrow m}P(n)-\sum_n
T_{m\rightarrow n}P(m),
\end{equation}
\noindent where $T_{n\rightarrow m}$ represents the
$\emph{transition amplitude}$
 or $\emph{propagator}$ from a microstate $n$ to a microstate $m$ and $P(m)$
 is the probability to find the system in state $m$. Considering a d-dimensional $\emph{lattice}$ $\Bbb{L}$ with lattice constant $l$,
 the microstates correspond to the occupation numbers $\{N_i\}=\{N_1,N_2,...\}$ at each lattice site $i$.
\\ \\
The chemical reaction of pairs of species $A$ to form a product of
species $C$ on a particle or droplet , $A + A \longrightarrow C$, is
modelled by the following master equation

\begin{eqnarray}
\nonumber
&&\frac{dP(\{N_A\},\{N_C\};t)}{dt}=j_A\sum_i \big(P(...,N_{A_i}-1,...,\{N_C\};t)-P(\{N_A\},\{N_C\};t)\big)\\
\nonumber
&&\phantom{dP(\{N_A\})}+j_C\sum_i \big(P(\{N_A\}...,N_{C_i}-1,...;t)-P(\{N_A\},\{N_C\};t)\big)\\
\nonumber
&&\phantom{dP(\{N_A\})}+\frac{\kappa}{V}\sum_i \big((N_{A_i}+2)(N_{A_i}+1)P(...,N_{A_i}+2,...N_{C_i}-1,...;t)\\
\nonumber
&&\phantom{\frac{dP(\{N\};t)}{dt}=j_A\sum_i \big(P}-N_{A_i}(N_{A_i}-1)P(\{N_A\},\{N_C\};t)\big)\\
\nonumber
&&\phantom{dP(\{N_A\})}+\lambda_A\sum_i \big((N_{A_i}+1)P(...,N_{A_i}+1,...,\{N_C\};t)-N_{A_i}P(\{N_A\},\{N_C\};t)\big)\\
\nonumber
&&\phantom{dP(\{N_A\})}+\lambda_C\sum_i \big((N_{C_i}+1)P(\{N_A\},...,N_{C_i}+1,...;t)-N_{C_i}P(\{N_A\},\{N_C\};t)\big)\\
\nonumber &&\phantom{dP(\{N_A\})}+D_A \sum_{\langle ij \rangle}
\big((N_{A_i}+1)P(...,N_{A_i}+1,N_{A_j}-1,...,\{N_C\};t)-N_{A_i}P(\{N_A\}\{N_C\};t) \\
\nonumber
&&\phantom{dP(\{N_A\})} + (N_{A_j}+1)P(...,N_{A_i}-1,N_{A_j}+1,...,\{N_C\};t)-N_{A_j}P(\{N_A\},\{N_C\};t)\big)\\
\nonumber &&\phantom{dP(\{N_A\})}+D_C \sum_{\langle ij \rangle}
\big((N_{C_i}+1)P(\{N_A\},...,N_{C_i}+1,N_{C_j}-1,...;t)-N_{C_i}P(\{N_A\}\{N_C\};t) \\
\nonumber
&&\phantom{dP(\{N_A\})} + (N_{C_j}+1)P(\{N_A\},...,N_{C_i}-1,N_{C_j}+1,...;t)-N_{C_j}P(\{N_A\},\{N_C\};t)\big),\\
&&\label{mAAC}
\end{eqnarray}
\noindent This equation describes the evolution of the probability
distribution $P(\{N_A\}, \{N_C\};t)$ for the total number of
adsorbed molecules $\{N_A\}$ of species $A$ and for the number of
reaction products $\{N_C\}$ of species $C$. The symbols
$N_{A_i,C_i}$ denote the numbers of $A$ or $C$ molecules at lattice
site $i$, respectively. The constants $j_{A,C}$, $\kappa$ and
$\lambda_{A,C}$ are rate coefficients, $D_{A,C}$ is the diffusion
constant and $V$ stands for the volume of the droplet. The rate
coefficient $j_{A,C}$ is called $\emph{source rate}$, the rate
coefficient $\lambda_{A,C}$ is called $\emph{evaporation rate}$ and
$\kappa$ is known as the $\emph{reaction rate}$.
\\ \\
In our model, the chemical reaction is taking place on a
d-dimensional lattice,
allowing for multiple occupancy on each site. This configuration is also called $\emph{bosonic representation}$.\\
The changes in population which we consider are caused by:
\begin{itemize}
\item[(a)] absorption of a molecule of species A from outside the grain particle (first line in the above equation),
and absorption of a molecule of species C from outside the grain
(second line in the above equation),
\item[(b)] binary reaction on the surface of the grain (third and fourth line in the above equation),
\item[(c)] evaporation of a molecule of species A from the grain (fifth line in the above equation),
and evaporation of a molecule of species C from the grain particle
(sixth line in the above equation),
\item[(d)] particle hopping of a molecule of species A from site $i$ to site $j$ (seventh line in the above equation),
and particle hopping of a molecule of species C from site $i$ to
site $j$ (eighth line in the above equation),
\item[(e)] particle hopping of a molecule of species A from site $j$ to site $i$ (ninth line in the above equation),
and particle hopping of a molecule of species C from site $j$ to
site $i$ (last line in the above equation).
\end{itemize}
 The summation in the
fourth and fifth line of the above equation \eqref{mAAC} is taken
over nearest neighbour sites only. The factors
$(N_{A,C}+2),(N_{A,C}+1),N_{A,C}, (N_{A,C}-1)$ describe the number
of ways of choosing the particles involved in the considered
process. In the continuum limit the particle hopping from one site
to another corresponds
to the diffusion of the particles.\\
The initial condition is chosen corresponding to a Poissonian
distribution on each site
\begin{equation}
\label{Poisson} P(\{N_A\},\{N_C\};t=0)=e^{-n_A(0)-n_C(0)}\prod_i
\frac{n_A(0)^{N_{A_i}}n_C(0)^{N_{C_i}}}{N_{A_i}!N_{C_i}!},
\end{equation}
\noindent where $n_{A,C}(0)$ is the average occupation number per
lattice site for the $A$ or $C$ particles respectively.
\\ \\
In the next step we will apply the methods of second quantisation
\cite{Doi;76I, Doi;76II}. We will rewrite the master equation as a
Schr\"odinger-like equation for a many-body wave function. This
approach can be justified by noting that, first of all, the master
equation is a differential equation of first order with respect to
time. The second reason to suggest the treatment of the master
equation according
to the second quantisation is that the master equation is linear in the probability.\\ \\
In order to simplify the notation we will suppress the dependence on
space coordinates $\mathbf{x}=(x_1,x_2,...,x_d)$. We will be working
in an appropriate space, the $\emph{Fock space}$. A Fock space
$\mathcal{F}_\nu(\mathcal{H})$ is a Hilbert space made from the
direct sum of tensor products of single-particle Hilbert spaces
$\mathcal{H}$
\begin{equation}
\mathcal{F}_\nu(\mathcal{H})=\bigoplus_{n=0}^\infty S_\nu
\mathcal{H}^{\otimes n},
\end{equation}
\noindent with $S_\nu$ a symmetrising (for the case of bosons) or
antisymmetrising (for the case of fermions) operator. The Fock space
is constructed by introducing the following operators at each
lattice site $i$
\begin{eqnarray*}
&&\co_i,\ i\in \Bbb{L}: \text{creation operator},\\
&&\ao_i,\ i\in \Bbb{L}: \text{annihilation operator},\\
\end{eqnarray*}
\noindent which satisfy the commutation relationships
\begin{equation}
\frac{1}{2}[\co_i,\ao_j]:= \frac{1}{2}(\co_i \ao_j - \ao_i
\co_j)=\delta_{ij}.
\end{equation}
\noindent
The $\emph{vacuum state}$ $| \{ 0 \} \rangle$ is defined by
\begin{equation}
\ao_i |\{0\}\rangle =|\{0\}\rangle \ \ \ \forall i \in \Bbb{L},
\end{equation}
\noindent with
\begin{equation}
|\{0\}\rangle:= \bigotimes_j |0_j\rangle,
\end{equation}
\noindent where $|0\rangle_j$ denotes the vacuum state in a
single-particle Hilbert space.
\\ \\
The master equation \eqref{mAAC} is equivalent to the
Schr\"odinger-like equation ---a Schr\"odinger equation with
imaginary time---
\begin{equation}
\frac{d}{dt}|\Psi \rangle_{A+A\rightarrow
C}=-\mathtt{H}_{A+A\rightarrow C}[\co_{A_i}, \ao_{A_j}, \co_{C_k},
\ao_{C_l}]|\Psi \rangle_{A+A\rightarrow C} \label{SlAAC}
\end{equation}
\noindent with the many-body wave function
\begin{equation}
\label{manybody} |\Psi \rangle _{A+A\rightarrow C}:= \sum_{\{N_A\},
\{N_C\}} P(\{N_A\},\{N_C\};t) \prod_i (\co_{A_i})^{N_{A_i}}
(\co_{C_i})^{N_{C_i}}|\{0\}\rangle,
\end{equation}
\noindent and the Hamiltonian operator
\begin{eqnarray}
\nonumber
&&\mathtt{H}_{A+A\rightarrow C}[\co_{A_i}, \ao_{A_j}, \co_{C_k}, \ao_{C_l}]=\sum_{M \in \{A,C\}}\sum_i(\co_{M_i}-\mathtt{1}_i)(j_M\mathtt{1}_i-\lambda_M \ao_{M_i})\\
\nonumber
&&\phantom{\mathtt{H}_{A+A\rightarrow C}[\co_{M_i}, \ao_{M_j};_{M \in \{A,C\}}]=}-\frac{\kappa}{V}\sum_i \big(\co_{C_i}-\co_{A_i}^2\big)\ao_{A_i}^2 \\
\label{HamAAC} &&\phantom{\mathtt{H}_{A+A\rightarrow C}[\co_{M_i},
\ao_{M_j};_{M \in \{A,C\}}]=}+ \sum_{M \in \{A,C\}}D_M\sum_{\langle
ij \rangle}(\co_{M_i} - \co_{M_j})(\ao_{M_i} - \ao_{M_j}).
\end{eqnarray}
\noindent For verification of the above statement one has to insert
the states $\prod_i (\co_{A_i})^{N_{A_i}}
(\co_{C_i})^{N_{C_i}}|\{0\}\rangle$ on both sides of the master
equation --- equation \eqref{mAAC}--- and sum over the set of all
occupation numbers $\{N_A\}$ and $\{N_C\}$. In general, the
time-evolution operator $\mathtt{H}$ is not necessarily Hermitian.
For the timebeing let us suppress the subindices that identify the
particle type. The form of the many-body wave function
\eqref{manybody} can be made plausible when considering the
\emph{state vector} $|N_i \rangle$ at site $i \in \Bbb{L}$, namely
\begin{equation}
|N_i \rangle = \co_i^{N_i}|0 \rangle.
\end{equation}
\noindent It holds that
\begin{equation}
 \ao_i |N_i\rangle =N_i |N_i-1 \rangle, \ \ \ \co_i |N_i
\rangle = |N_i +1 \rangle.
\end{equation}
\noindent

\subsection{Expectation Values Of Observables}

We are interested in obtaining the expectation values for various
observables, especially the average number density. The expectation
values of observables $\mathtt{O}$ are given by
\begin{equation}
\label{expval} \langle \mathtt{O} \rangle
:=\sum_{\{N_i\}}\mathtt{O}(\{N_i\})P(\{N_i\};t).
\end{equation}
\noindent We want the expectation values of the observables
---diagonal in the occupation number basis--- to be linear in the
probabilities. After some straightforward computation ---see for
example \cite{TaeHowVol;05}--- one can see that the above expression
is equivalent to
\begin{equation}
\langle \mathtt{O} \rangle = \langle \{\mathtt{P}\} |\mathtt{O}|
\Psi(t) \rangle,
\end{equation}
\noindent where
\begin{equation}
 \langle \{\mathtt{P}\} |:=\langle \{0\}|e^{\sum_j\ao_j}:
\text{\emph{projection state}}.
\end{equation}
\noindent The projection state obeys the relation
\begin{equation}
\langle \{\mathtt{P}\} |\{ 0\}\rangle =1.
\end{equation}
\noindent By definition, the projection state is a left eigenstate
of all creation operators with unit eigenvalue
\begin{equation}
 \langle \{\mathtt{P}\} | \co_i = \langle \{\mathtt{P}\}|,
\ \ \ \forall i \in \Bbb{L}.
\end{equation}
\noindent Furthermore, $\langle \{\mathtt{P}\} | \Psi(t) \rangle
=1$. Conservation of probability requires that $\langle \{
\mathtt{P}\} | \mathtt{H}=\langle\{ 0\}|$.
\\ \\
We break the time interval $[t_0,t_T]$ into T short slices of
duration $\Delta t=\frac{t_T-t_0}{T}$. At each time slice we insert
a complete set of coherent states ---see, for example,
\cite{KlaSka;85}. Coherent states, $|\mathtt{C}_i(t) \rangle$, are
right eigenstates of the annihilation operator
\begin{equation}
 \ao_i |\mathtt{C}_i(t) \rangle = \Phi_i(t)
|\mathtt{C}_i(t) \rangle, \ \ \ i \in \Bbb{L}
\end{equation}
\noindent where the eigenvalue $\Phi_i$ is a complex function. The
duals $\langle \mathtt{C}_i(t)|$ are left eigenstates of the
creation operator
\begin{equation}
\langle \mathtt{C}_i(t)| \co_i = \langle \mathtt{C}_i(t)|
\Phi_i^{\ast}(t), \ \ \ i \in \Bbb{L}.
\end{equation}
\noindent We have
\begin{eqnarray}
\nonumber
&&|\mathtt{C}_i(t)\rangle := e^{-\frac{1}{2}| \Phi_i(t)|^2 + \Phi_i(t)\co_i}|0\rangle,\\
&&\langle \mathtt{C}_i(t)|: = \langle 0|e^{-\frac{1}{2}|
\Phi_i^{\ast}(t)|^2 +\Phi_i^\ast(t)\ao_i}.
\end{eqnarray}
\noindent The coherent states are over-complete. Still, we can use
them to create the identity
\begin{equation}
\mathtt{1}=\frac{1}{\pi}\int
d[\emph{Re}(\Phi_i)]d[\emph{Im}(\Phi_i)] |\mathtt{C}_i(t)\rangle
\langle \mathtt{C}_i(t)|,
\end{equation}
\noindent for a single lattice site $i \in \Bbb{L}$, and for
multiple lattice sites accordingly
\begin{equation}
 \label{id} \mathtt{1}=\int \prod_i
\bigg(\frac{1}{\pi}d[\emph{Re}(\Phi_i)]d[\emph{Im}(\Phi_i)] \bigg)
|\{\mathtt{C}\}\rangle \langle \{\mathtt{C}\}|,
\end{equation}
\noindent with $|\{\mathtt{C}\}\rangle=\bigotimes _j |\mathtt{C}_j
\rangle$. Let us recall the formula for the expectation values of
observables
\begin{equation}
\label{obs} \langle \mathtt{O} \rangle = \langle \{\mathtt{P}
\}|\mathtt{O}| \Psi(t) \rangle=\langle \{0\}| e^{\sum_i
\ao_i}\mathtt{O}e^{-\mathtt{H}t}| \Psi(0) \rangle,
\end{equation}
\noindent where the initial many-body wave function takes the form
---see equations \eqref{Poisson} and \eqref{manybody}---
\begin{equation}
| \Psi(0) \rangle:= e^{n(0)\big(\sum_i\co_i-1\big)}|\{0\}\rangle
\end{equation}
\noindent We observe the following proportionalities
\begin{eqnarray}
\nonumber
&&\langle \{ \mathtt{P}\}| \propto \langle \{\mathtt{C}(t)\}|_{\Phi^{\ast}_j=1}=\langle \{0\}| e^{-\frac{1}{2}+\sum_i\ao_i},\\
&&| \Psi(t=0) \rangle \propto |\{\mathtt{C}(t)\}
\rangle_{\Phi_j=n(0)}=e^{-\frac{1}{2}|n(0)|^2 + n(0)\sum_i
\co_i}|\{0\} \rangle,
\end{eqnarray}
\noindent for all admissible values of $j$. Therefore, one can
recast the equation for the expectation values \eqref{obs} into
\begin{equation}
\label{obsnew} \langle \mathtt{O} \rangle \propto \langle
\{\mathtt{C}_1(t)\}|\mathtt{O} e^{-\mathtt{H}t}|\{
\mathtt{C}_{n(0)}(t)\}\rangle  ,
\end{equation}
\noindent where
\begin{eqnarray}
\nonumber
&&\langle \{\mathtt{C}_1(t)\}|:=\langle \{\mathtt{C}(t)\}|_{\Phi^{\ast}_j=1},\\
&&|\{ \mathtt{C}_{n(0)}(t)\}\rangle
:=|\{\mathtt{C}(t)\}\rangle_{\Phi_j=n(0)},
\end{eqnarray}
\noindent for all admissible values of $j$. According to the
breakage of the time interval into time slices of small duration we
rewrite the expression
\begin{equation}
e^{-\mathtt{H}t}=e^{-\mathtt{H}\Delta t}e^{-\mathtt{H}\Delta
t}......(\text{T times})
\end{equation}
\noindent occuring in the equation for the expectation values
\eqref{obsnew} and insert the identity as defined in \eqref{id}
between each factor. Then the discrete version of the expectation
values of operators $\mathtt{O}$ reads
\begin{eqnarray}
\nonumber
&&\langle \mathtt{O} \rangle _{discrete}\propto \int \bigg(\prod_{i,\tau} d[\emph{Re}(\Phi_{i,\tau})]d[\emph{Im}(\Phi_{i,\tau})]\bigg)\langle \{\mathtt{C}_1\}|\mathtt{O}|\{\mathtt{C}_\tau \}\rangle \times ...\\
&&\phantom{\langle \mathtt{O} \rangle _{discrete}=}....\times
\bigg(\prod^T_{\tau=\Delta t}\langle \{\mathtt{C}_\tau\}
|e^{-\mathtt{H}\Delta t}|\{\mathtt{C}_{\tau-\Delta t} \}\rangle
\bigg) \langle \{\mathtt{C}_{\tau=0}\}|\{\mathtt{C}_{n(0)}
\}\rangle,
\end{eqnarray}
\noindent where we have labeled each time slice by a time index
$\tau \in [0,\Delta t, 2\Delta t, ..., T]$. The normalisation
constant has to be determined lateron. The consideration of the
lattice expectation value is  not sufficient if one is interested in
long wavelength properties. In the formal continuum limit, we obtain
\begin{equation}
\langle \mathtt{O} \rangle_{continous}\equiv \langle \mathtt{O}
\rangle = \lim_{\Delta t \longrightarrow 0} \langle \mathtt{O}
\rangle_{discrete}.
\end{equation}
\noindent  Next, we expand the exponential function for small
$\Delta t$, neglect higher order terms in $\Delta t $ and recast the
continuous average. When interested in inclusive probabilities
---e.g. the average number of particles at a given lattice site
irrespective of the number of particles elsewhere--- it is
convenient to commute the factor of $e^{\sum_i\ao_i}$ through the
operators $\mathtt{O}$ and $\mathtt{H}$ in $\langle
 \mathtt{O} \rangle$. This has the effect of shifting $\co \rightarrow \co +1$ using $e^{\ao} \co=(\co+1)e^{\ao}$.
 The operators are now normal ordered. It is valid that the operator
 $\mathtt{O}$ and its normal ordered counterpart have the same
 expectation value if all creation operators occuring in the normal ordered operator are replaced by unity
 ---see for example \cite{TaeHowVol;05}. In particular, the density
 operator $\co \ao$ reduces to the annihilation operator $\ao$. Therefore, in the continuum limit
  the average particle density of the $A$
particles is given by the stochastic average of the complex
eigenvalues of the coherent state vectors under the annihilation
operator, that is one chooses the operator $\mathtt{O}$ to be
$\Phi_A(\mathbf{x}, t)$.\\ \\
We take the continuum limit ---the dimensions of the constants are
chosen by looking at the discrete Hamiltonian operator
\eqref{HamAAC}--- via $\sum_i \longrightarrow \int l^{-D} \ d^Dx$,
$\Phi_{A_i,C_j}(t) \rightarrow \Phi_{A,C}(\mathbf{x},t)l^D$,
$\Phi^{\ast}_{A_i,C_j}(t) \rightarrow
\Phi^\ast_{A,C}(\mathbf{x},t)$, $D_{A,C} \rightarrow
\bar{D}_{A,C}l^{-2}$, $\kappa/V \rightarrow \bar{\kappa}l^{-D}$,
$\lambda_{A,C} \rightarrow \bar{\lambda}_{A,C}$, $j_{A,C}
\rightarrow \bar{j}_{A,C}l^D$ and finally $n_{A,C}(0) \rightarrow
\bar{n}_{A,C}(0)l^D$, where the newly introduced constants have the
following dimension properties $[\Phi_{A,C}(\mathbf{x},t)]=m^{-D}$,
$[\bar{D}_{A,C}]=m^2s^{-1}$, $[\bar{\kappa}]=m^Ds^{-1}$,
$[\bar{j}_{A,C}]=m^{-D}s^{-1}$ and $[\bar{n}_{A,C}(0)]=m^{-D}$ in
Standard International units. The objects
$\Phi^\ast_{A,C}(\mathbf{x},t)$ and $\bar{\lambda}_{A,C}$ are
dimensionless. Notice that now $\Phi_{A,C}(\mathbf{x},t)$ scales
like a density.
\\ \\
In the continuum limit, the average particle density in the
stochastic model is then given by
\begin{eqnarray}
\nonumber && \langle \Phi_A(\mathbf{x},t)\rangle:= \langle\{
0\}|\Phi_A(\mathbf{x},t)
e^{-\tilde{S}_{A+A\rightarrow C}[\{\Phi_{A,C}\},\{\tilde{\Phi}_{A,C}\}]}|\{0\} \rangle\\
\nonumber &&\phantom{n(x,t)\equiv}=\frac{\int D\Phi_A D\Phi_C
D\tilde{\Phi}_A D\tilde{\Phi}_C \Phi_A(\mathbf{x},t)
e^{-\tilde{S}_{A+A\rightarrow
C}[\{\Phi_{A,C}\},\{\tilde{\Phi}_{A,C}\}]}}
{\int D\Phi_A D\Phi_C D\tilde{\Phi}_A D\tilde{\Phi}_Ce^{-\tilde{S}_{A+A \rightarrow C}[\{\Phi_{A,C}\},\{\tilde{\Phi}_{A,C}\}]}},\\
&&\label{newexpval}
\end{eqnarray}
\noindent where $D$ denotes the $\emph{measure}$ of the functional
integral and $\tilde{\Phi}_{A,C}$ is the $\emph{shifted eigenvalue}$
of the dual of the coherent state under the creation operator
defined by $\tilde{\Phi}_{A,C}:=\Phi^{\ast}_{A,C}-1$. Accordingly,
all fields that incorporate shifted eigenvalues instead of the
original eigenvalues will be denoted by a twiddle in the sequel.
Note that the average \eqref{newexpval} is performed taking into
account the dynamics and the initial conditions ---for a more
detailed discussion see \cite{TaeHowVol;05}. We already have
incorporated the \emph{shifted initial state}
\begin{equation}
|\tilde{\Psi}(t=0)\rangle=e^{\int d^D x
(\bar{n}_A(0)\tilde{\Phi}_A(t=0)+\bar{n}_C(0)\tilde{\Phi}_C(t=0))}|\{0\}\rangle,
\end{equation}
\noindent in the \emph{shifted action} $\tilde{S}$. The symbol
$\tilde{S}_{A+A \rightarrow C}$ represents the $\emph{shifted
action}$ which is defined as follows
\begin{equation}
\tilde{S}_{A+A\rightarrow C}:=\int_0^{t_T} dt \int d^dx
\{\tilde{\Phi}_A \} \pd{\{ \Phi_A \}}{t} + \{\tilde{\Phi}_C \}
\pd{\{ \Phi_C \}}{t}+ \tilde{H}_{A+A\rightarrow C}[\{ \Phi_A \}, \{
\tilde{\Phi}_A \}, \{ \Phi_C \}, \{ \tilde{\Phi}_C \}],
\end{equation}
\noindent with the shifted Hamiltonian $\tilde{H}_{A+A \rightarrow
C}$. The shifted action for the chemical reaction
$A+A\longrightarrow C$ takes the form
\begin{eqnarray}
\nonumber &&\tilde{S}_{A+A\rightarrow
C}[\{\Phi_A\},\{\tilde{\Phi}_A\},
\{\Phi_C\},\{\tilde{\Phi}_C\}]=\int_{0}^{t_T} dt
\int d^Dx \bigg(\sum_{M \in \{A,C\}}\big(\tilde{\Phi}_M( \pd{}{t}-\bar{D}_M\Delta)\Phi_M\\
\nonumber &&\phantom{\tilde{S}_{A+A\rightarrow
C}[\{\Phi\},\{\tilde{\Phi}\}]=\int_{0}^{t_T}}-
\tilde{\Phi}_M(\bar{j}_M-\bar{\lambda}_M
\Phi_M)\big)+\bar{\kappa}(2\tilde{\Phi}_A+\tilde{\Phi}_A^2
-\tilde{\Phi}_C)\Phi_A^2\bigg)\\
\label{shiftedaction} &&\phantom{\tilde{S}_{A+A\rightarrow
C}[\{\Phi\},\{\tilde{\Phi}\}]=\int_{0}^{t_T}} -\int d^D
x\big(\bar{n}_A(0)\tilde{\Phi}_A(t=0)+\bar{n}_C(0)\tilde{\Phi}_C(t=0)\big).
\end{eqnarray}
\noindent We want to untangle the quadratic term $\tilde{\Phi}_A^2$.
A linear expression in $\tilde{\Phi}_A$ can be obtained by means of
a Gaussian transformation
\begin{equation}
e^{-\bar{\kappa}\int_{0}^{t_T} dt\int
d^Dx\tilde{\Phi}_A^2(\mathbf{x},t)\Phi_A^2(\mathbf{x},t)} \propto
\int D\eta \ \mathcal{P}[\eta]_{A+A\rightarrow
C}e^{\mbox{i}\sqrt{2\bar{\kappa}}\int_{0}^{t_T} dt\int d^Dx
\tilde{\Phi}_A(\mathbf{x},t)\Phi_A(\mathbf{x},t)\eta(\mathbf{x},t)},
\end{equation}
\noindent where $\mathcal{P}[\eta]_{A+A\rightarrow C}$ is the
probability distribution for a white noise $\eta(\mathbf{x},t)$. The
Gaussian distribution reads
\begin{equation}
\label{GdAAC} \mathcal{P}[\eta]_{A+A\rightarrow
C}=e^{-\frac{1}{2}\int_{0}^{t_T} dt \int d^Dx \eta^2(\mathbf{x},t)}.
\end{equation}
\noindent Now the shifted action $\tilde{S}$ is linear in
$\tilde{\Phi}_A$ and one can easily integrate out over
$\tilde{\Phi}_A(\mathbf{x},t)$ and $\tilde{\Phi}_C(\mathbf{x},t)$ in
\eqref{newexpval}. One obtains
\begin{eqnarray}
\nonumber &&\langle\mathtt{O}[\Phi_A,\Phi_C] \rangle \propto \int
D\Phi_A D\Phi_C
D\eta \ \mathtt{O}[\Phi_A,\Phi_C] \delta [\mathcal{F}_A]\delta [\mathcal{F}_C] P[\eta]_{A+A\rightarrow C}\\
\nonumber &&\phantom{\langle \langle\mathtt{O}[\Phi_A,\Phi_C]
\rangle \rangle } \propto \int D\eta \
\mathtt{O}[\bar{\Phi}_A[\eta(\mathbf{x},t),\mathbf{x},t],\bar{\Phi}_C[\eta(\mathbf{x},t),\mathbf{x},t]]
P[\eta]_{A+A\rightarrow C},\\
\label{average1}
\end{eqnarray}
\noindent where $\delta[\mathcal{F}_{A,C}]$ is a functional Dirac
delta distribution. In its generalised Fourier representation it is
defined by
\begin{equation} \delta [\mathcal{F}]:= constant
\int d \lambda(y) e^{\int dy \lambda(y)
\mathcal{F}[\mathbf{z}(y),y]},
\end{equation}
\noindent with $\mathbf{z}(y)$ being a multicomponent field
satisfying the constraint
\begin{equation}
\mathcal{F}[\mathbf{z}(y),y]=0.
\end{equation}
\noindent Accordingly, the functions
$\bar{\Phi}_A[\eta(\mathbf{x},t),\mathbf{x},t]$ and
$\bar{\Phi}_C[\eta(\mathbf{x},t),\mathbf{x},t]$ satisfy the
constraints
\begin{eqnarray}
\nonumber
&&\mathcal{F}_A[\bar{\Phi}_A(\mathbf{x},t),\mathbf{x},t]\equiv -\pd{\bar{\Phi}_A(\mathbf{x},t)}{t}+\bar{D}_A\Delta \bar{\Phi}_A(\mathbf{x},t)  -2\bar{\kappa}\bar{\Phi}_A^2(\mathbf{x},t)-\bar{\lambda}_A\bar{\Phi}_A(\mathbf{x},t)\\
\label{Ll2AAC}
&&\phantom{\mathcal{F}[\Phi_A(\mathbf{x},t),\mathbf{x},t]=} +\bar{j}_A +\mbox{i}\sqrt{2\bar{\kappa}}\bar{\Phi}_A(\mathbf{x},t)\eta(\mathbf{x},t)=0,\\
\nonumber
&&\mathcal{F}_C[\bar{\Phi}_C(\mathbf{x},t),\mathbf{x},t] \equiv -\pd{\bar{\Phi}_C(\mathbf{x},t)}{t}+\bar{D}_C\Delta \bar{\Phi}_C(\mathbf{x},t)  +\bar{\kappa}\bar{\Phi}_A^2(\mathbf{x},t)-\bar{\lambda}_C\bar{\Phi}_C+\bar{j}_C=0.\\
&&\label{LlAAC}
\end{eqnarray}
\noindent It follows from equation \eqref{average1} that the average
particle density for the A respectively C molecules is now given by
\begin{equation}
\label{average} \langle \Phi_{A,C}(\mathbf{x},t) \rangle = \int
D\eta \ \bar{\Phi}_{A,C}[\eta(\mathbf{x},t),\mathbf{x},t]
e^{-\frac{1}{2}\int_{0}^{t_T} dt \int d^Dx \eta^2(\mathbf{x},t)}.
\end{equation}
\noindent The stochastic noise $\eta$  has zero mean value and a
correlation given by
\begin{equation}
\label{corr} \langle \eta(\mathbf{x},t)\eta(\mathbf{x}', t')
\rangle_{\mathcal{P}[\eta]_{A+A\rightarrow C}}
=\delta^{(D)}(\mathbf{x}-\mathbf{x}')\delta(t-t').
\end{equation}
\noindent This is obvious when considering the Gaussian distribution
\eqref{GdAAC}. Note that the above average is no longer taken over
the initial conditions. \\
The constraint equation \eqref{Ll2AAC} is an inhomogeneous partial
stochastic differential equation with additive noise for a complex
fluctuating unknown field in the It$\hat{\text{o}}$ calculus. It
resembles the deterministic partial differential equation that
describes the evolution of the mean particle density in the
classical theory. But despite the temptation for an intuitive
interpretation it is very important to keep in mind that in equation
\eqref{Ll2AAC} we are confronted with a complex fluctuating quantity
that has as such no physical interpretation. Only if the path
integral average (PIA) \eqref{average} of a solution to
\eqref{Ll2AAC} or \eqref{LlAAC} is taken over all possible
realisations of the stochastic noise that appears in the constraint
equation can one interpret the outcome of this computation as a mean
particle density.


\section{Case A: Vanishing Source Rate}

In the remainder of this paper, let us concentrate on a single
spatial site model. We will now compare the results in the
stochastic model to the observations made in the traditional
approach. The classical equation for the evolution of the mean
particle density in the single spatial site model reads
\begin{equation}
\label{classicalA}
 \frac{d}{dt} n_A(t)  +
2\bar{\kappa} n_A^2(t)+ \bar{\lambda}_A n_A(t)=0
\end{equation}
\noindent where $n_A(t)$ denotes the mean particle density of the
$A$ molecules in the mean field approximation. The classical
evolution equation \eqref{classicalA} is solved by
\begin{equation}
\label{classA} n_A(t)= \frac{\bar{\lambda}_A}{2\bar{\kappa}}
\frac{1}{-1 + e^{\bar{\lambda}_A
t}\big(1+\frac{\bar{\lambda}_A}{2\bar{\kappa}}n_A(0)\big)}.
\end{equation}
\noindent The stochastic constraint equation for the complex
fluctuating field $\bar{\Phi}_A(t)$ in zero spatial dimensions with
vanishing source rate takes the form
\begin{equation}
\label{consA} \frac{d}{dt}\bar{\Phi}_A(t)
+2\bar{\kappa}\bar{\Phi}_A^2(t)+\bar{\lambda}_A\bar{\Phi}_A(t)
 -\mbox{i}\sqrt{2\bar{\kappa}}\bar{\Phi}_A(t)\eta(t)=0.
\end{equation}
\noindent The traditional equation for the average particle density
of the A molecules in the mean field approach \eqref{classicalA} and
the stochastic constraint equation associated with the A molecules
\eqref{consA} resemble each other at first sight. But as mentioned
before the solution of the stochastic differential equation
\eqref{consA} is a complex, fluctuating field that can only be
interpreted as an average particle density after it has been
averaged in the sense of equation \eqref{average}. For vanishing
source rate it is fairly easy to find an analytic solution of
equation \eqref{consA}. The stochastic constraint equation
\eqref{consA}
---because of the continuous but not smooth nature of a stochastic process--- has to be
understood in terms of a stochastic integral equation
\begin{equation}
 \bar{\Phi}_A(t)-\bar{\Phi}_A(t=0)=\int_0^t ds\
a(\bar{\Phi}_A(s),s)+\int_0^tds \ b(\bar{\Phi}_A(s),s)\eta(s),
\end{equation}
\noindent where
\begin{eqnarray}
\nonumber
&&a(\bar{\Phi}_A(t),t)=\bar{j}_A-\bar{\lambda}_A\bar{\Phi}_A(t)-2\bar{\kappa}\bar{\Phi}_A^2(t):\ \text{\emph{drift coefficient}},\\
&&b(\bar{\Phi}_A(t),t)=\mbox{i}\sqrt{2\bar{\kappa}}\bar{\Phi}_A(t):
\ \text{\emph{diffusion coefficient}}.
\end{eqnarray}
\noindent The stochastic noise $\eta(t)$ is rewritten in terms of
the Wiener process $W(t)$
\begin{equation}
\eta(t)dt=dW(t).
\end{equation}
\noindent For vanishing source rate $\bar{j}_A$ the stochastic
constraint equation for the $A$ particle density \eqref{Ll2AAC}
reduces to the following equation
\begin{equation}
\label{verhulst}
d\bar{\Phi}_A(t)=\big(-2\bar{\kappa}\bar{\Phi}_A^2(t)-\bar{\lambda}_A\bar{\Phi}_A(t)\big)dt
+\mbox{i}\sqrt{2\bar{\kappa}}\bar{\Phi}_A(t)dW(t).
\end{equation}
\noindent The above equation is a nonlinear reducible stochastic
differential equation with polynomial drift of degree two in the
It$\hat{\text{o}}$ picture. In contrary to a Stratonovich stochastic
differential equation an It$\hat{\text{o}}$ stochastic differential
equation can not be solved directly by methods of classical
calculus\footnote{Sample paths of a Wiener process are ---with
reasonable certainty--- neither differentiable nor of bounded
variation. As a consequence one is left with different
interpretations of stochastic equations, namely the
It$\hat{\text{o}}$ and the Stratonovich interpretation. For a
further reading we refer to \cite{KloPla;92}.}. For an analytical
solution of an It$\hat{\text{o}}$ stochastic differential equation
one has to use a modified version of the drift coefficient
\begin{equation}
 a(\bar{\Phi}_A(t),t) \longrightarrow a(\bar{\Phi}_A(t),t)
- \frac{1}{2} b(\bar{\Phi}_A(t),t) \frac{\delta}{\delta
\bar{\Phi}_A(t)}b(\bar{\Phi}_A(t),t),
\end{equation}
\noindent where the derivative in the last term is a functional
derivative. Equation \eqref{verhulst} is a stochastic version of a
Verhulst-like equation
---see \cite{KloPla;92}. It can be reduced to a linear stochastic
differential equation with multiplicative noise.
We obtain the solution to the first stochastic constraint
equation \eqref{Ll2AAC} for vanishing source rate, namely
\begin{equation}
\label{solA0} \bar{\Phi}_A(t)=\frac{\bar{\Phi}_A(0)e^{(\bar{\kappa}
- \bar{\lambda}_A)t +\mbox{i}
\sqrt{2\bar{\kappa}}W_A(t)}}{1+2\bar{\kappa} \bar{\Phi}_A(0)\int_0^t
e^{(\bar{\kappa} - \bar{\lambda}_A)s +
\mbox{i}\sqrt{2\bar{\kappa}}W_A(s)}ds}.
\end{equation}
\noindent Inserting the above solution into the path integral
average \eqref{average} one obtains the average particle density for
the $A$ molecules in the stochastic picture.
\\ \\
The stochastic constraint equation for the reaction product, the $C$
particles, in zero dimensions looks ---in its form--- identical to
the classical evolution equation for the mean density of $C$
particles
\begin{equation}
\label{consC} \frac{d}{dt} n_C(t) + \bar{\lambda}_C n_C(t) -
\bar{\kappa} n_A^2(t) - \bar{j}_C=0
\end{equation}
\noindent In the single spatial site model, the full solution of the
second constraint equation
---simply take $\bar{\Phi}_C(t)$ instead of $n_C(t)$ in the above
equation \eqref{consC}--- can be obtained even for non-vanishing
source rate and reads
\begin{equation}
\label{solC0} \bar{\Phi}_C(t)=\bigg(\int_0^t e^{\bar{\lambda}_C s}
\big(\bar{\kappa} \bar{\Phi}_A^2(s) +\bar{j}_C \big) ds +
\bar{\Phi}_C(0)\bigg)e^{-\bar{\lambda}_C t}.
\end{equation}
\noindent The stochasticity of the above equation is hidden in the
first term employing the fluctuating solution $\bar{\Phi}_A(t)$ of
the first constraint equation \eqref{consA}.
\\ \\
Once the solutions to the stochastic constraint equations
\eqref{solA0} and \eqref{solC0} are known, one has to insert either
of the solutions into the path integral average \eqref{average} and
compute the path integral by means of a Monte Carlo calculation in
order to obtain the average particle density for the $A$ or $C$
particle population, respectively. Random samples are generated
according to the Gaussian probability distribution \eqref{GdAAC};
that is we generate Wiener processes. We estimate the path integral
\eqref{average} by summing a large number of solutions of the
constraint equations associated to the set of random samples
generated in the above sense and divide the sum by the number of
random samples. The Monte Carlo method displays a convergence of
$\frac{1}{\sqrt{N}}$ where $N$ is the number of random samples
---see \cite{Pre;92}.
\\ \\
 Instead of using the expressions of the
analytical solutions, equation \eqref{solA0} and equation
\eqref{solC0}, to generate solutions to the stochastic constraint
equations one can alternatively compute the paths directly from the
stochastic differential equations \eqref{Ll2AAC} and \eqref{LlAAC}.
The latter method turns out to be less time consuming. The
stochastic differential equation \eqref{Ll2AAC} in zero dimensions
can be converted into
\begin{equation}
\label{numerics}
X_{A,n+1}=X_{A,n}+(-\bar{\lambda}_AX_{A,n}-2\bar{\kappa}X_{A,n}^2)\Delta_n+\mbox{i}\sqrt{2\bar{\kappa}}X_{A,n}\Delta
W_n,
\end{equation}
\noindent where $X_{A,n}:=\Phi_A(t_n)$ in discretised time $t_n$ for
$n=0,..,N$, $\Delta_n:=t_{n+1}-t_n$ and $\Delta
W_n:=W_{t_{n+1}}-W_{t_n}$. The $\Delta W_n$ is generated by two
uniformly distributed independently random variables via the
Box-Muller transformation ---see, for example, \cite{KloPla;92}. The
numerical scheme \eqref{numerics} is called the Euler scheme and is
the most straightforward approach to undertake some numerical
investigations. Accordingly, the second constraint equation in the
single spatial site model and for vanishing source rate takes the
following form
\begin{equation}
X_{C,n+1}=X_{C,n}+(-\bar{\lambda}_CX_{C,n}+\bar{\kappa}X_{A,n}^2)\Delta_n.
\end{equation}
\noindent As stochastic differential equations are extremely
sensitive one has to convince oneself that the code is stable and
converging as it should be. Other schemes we used that might be more
accurate or stable than the Euler method are the Milstein scheme,
the simplified order 2.0 weak Taylor scheme, the implicit order 1.0
strong Runge-Kutta scheme, the predictor-corrector method of order 1
with modified trapezoidal method weak order 1.0 ---see
\cite{KloPla;92}.
\\ \\
For the numerical evaluation we employ values for the rate
coefficients that can be found in realistic physical set-ups.
Instead of using the coefficients introduced in the continuum limit,
namely $\bar{\lambda}_{A,C}$ and $\bar{\kappa}$, we employ the
traditional rate coefficients which we denote by $L_{A,C}$ and
$K_{A,C}$ and which have dimensions per unit time. Dimensional
analysis shows that using these new constants we are now calculating
an average particle population instead of an average particle
density ---this can be easily verified in equations \eqref{consA}
and \eqref{consC}.
\\ \\
The following plots were generated for the situation where two
hydrogen atoms react on the surface of an interstellar dust
particle. According to \cite{CasHasHer;98, StaCasHer;01} the
reaction rate takes the value of $K = 1.45 \times 10^5$ $s^{-1}$,
the evaporation rate for the hydrogen atoms $L_H= 1.88 \times
10^{-3}$ $s^{-1}$ and the evaporation rate for the reaction product
$L_{H_2}=6.9 \times 10^{-8}$ $s^{-1}$. As initial values we used
$\bar{\Phi}_A(0)=\bar{\Phi}_C(0)=6$.
\\ \\
In Figure 1 we generate the real part of one solution to the
stochastic constraint equation for the hydrogen atoms under the
above conditions. Figure 2 shows the imaginary part of the same
solution of the stochastic constraint equation for the reaction
partners. If one compares these plots to Figure 3 and Figure 4 where
the real and imaginary part of the path integral average over 1000
realisations of the white Gaussian noise for the $H$ atoms are given
one observes that the real part of the path integral average
smoothes out and the fluctuations in the imaginary part decrease in
intensity. For increasing number of paths employed in the path
integral average the imaginary part of the PIA tends to zero.
Therefore, it is safe to interpret the real part of the path
integral average as the average particle population.
\\ \\
Figures 5 and 6 show the real and imaginary part of a solution to
the second stochastic equation that constrains the reaction products
$H_2$. Again the fluctuations are smoothed out in Figure 7 and
Figure 8 when the path integral average for the diatomic hydrogen is
taken over 1000 realisations of the stochastic noise.
\\ \\
Together with Figure 7 one can interpret the path integral average
in Figure 3 in the following way: according to Figure 7 the chemical
reaction stops after certain transient processes. The average
population of the diatomic hydrogen is constant. The intuitive
physical reason for this is because all the potential reaction
partners $H$ have already been used to form the reaction product
$H_2$. On the other hand, from Figure 3 one sees how the plot for
the hydrogen atoms approaches asymptotically the value one half
which could be interpreted as a state consisting of either zero or
one particle. The value of $\Phi_A(t\rightarrow \infty)=1/2$ is the
lowest possible eigenvalue of the coherent states under the
annihilation operator once the system has reaches its equilibrium.
The rate coefficients $K$ and $L_{A,C}$ have to be understood in a
probabilistic sense, in a similar fashion as it is done with, say,
the mean life expectancy of a radioactive isotope.
For the specific values we used in our calculations the reaction
rate dominates over the evaporation rate. \\ \\The discussion above
can be compared with the results obtained from the solution to the
classical evolution equation \eqref{classA} which predicts an
asymptotic value $n_A(t) \longrightarrow 0$ as $t \longrightarrow
\infty$ for $L_A \neq 0$. That is, the classical model predicts the
extinction of all the reactants.

\begin{figure}[p]
\centering
\includegraphics[width=0.7\textwidth]{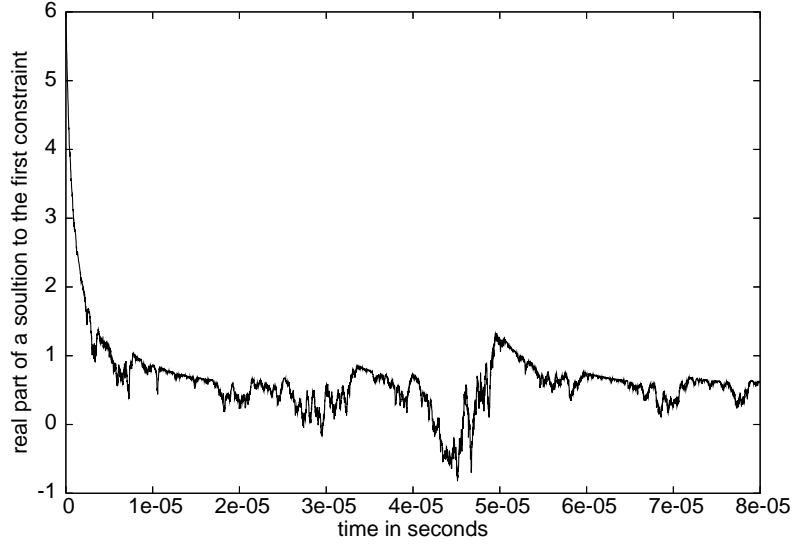}
\caption{The real part of one possible solution to the first
constraint equation \eqref{Ll2AAC} for the hydrogen atoms under
interstellar space conditions ($K=1.45 \times 10^5 s^{-1}$,
$L_H=1.88 \times 10^{-3}s^{-1}$) with vanishing source rates
($J_H=J_{H_2}=0 s^{-1}$).}
\end{figure}

\begin{figure}[p]
\centering
\includegraphics[width=0.7\textwidth]{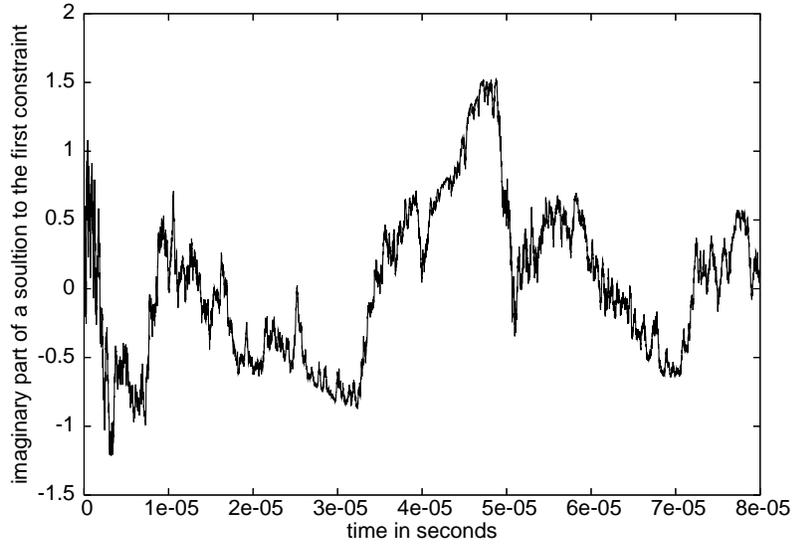}
\caption {The imaginary part of the solution to the first constraint
equation \eqref{Ll2AAC} for the same stochastic noise for the
reaction partners $H$ under interstellar space conditions ($K=1.45
\times 10^5 s^{-1}$, $L_H=1.88 \times 10^{-3}s^{-1}$) and for zero
source rate ($J_H=J_{H_2}=0 s^{-1}$).}
\end{figure}

\begin{figure}[p]
\centering
\includegraphics[width=0.7\textwidth]{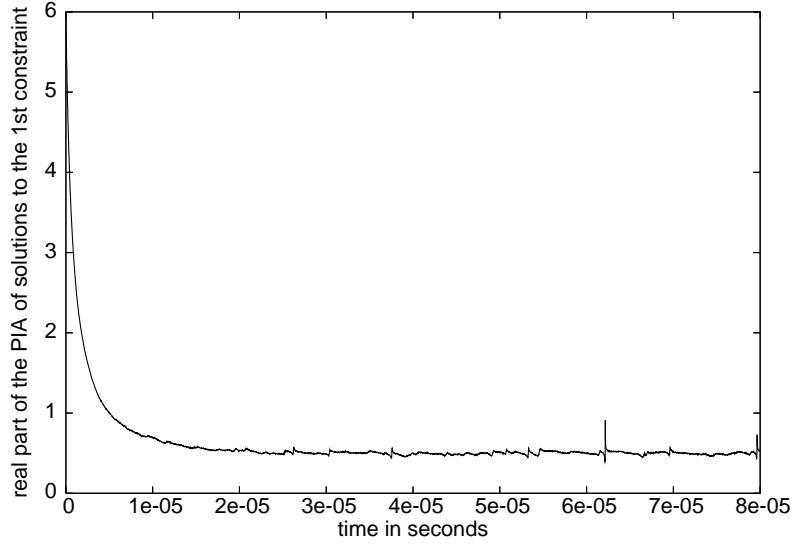}
\caption {The real part of the path integral average (PIA) of
solutions to the first constraint equation \eqref{Ll2AAC} with
vanishing source rate ($J_H=J_{H_2}=0 s^{-1}$) for the hydrogen
atoms under interstellar conditions ($K=1.45 \times 10^5 s^{-1}$,
$L_H=1.88 \times 10^{-3}s^{-1}$) over 1000 possible paths.}
\end{figure}

\begin{figure}[p]
\centering
\includegraphics[width=0.7\textwidth]{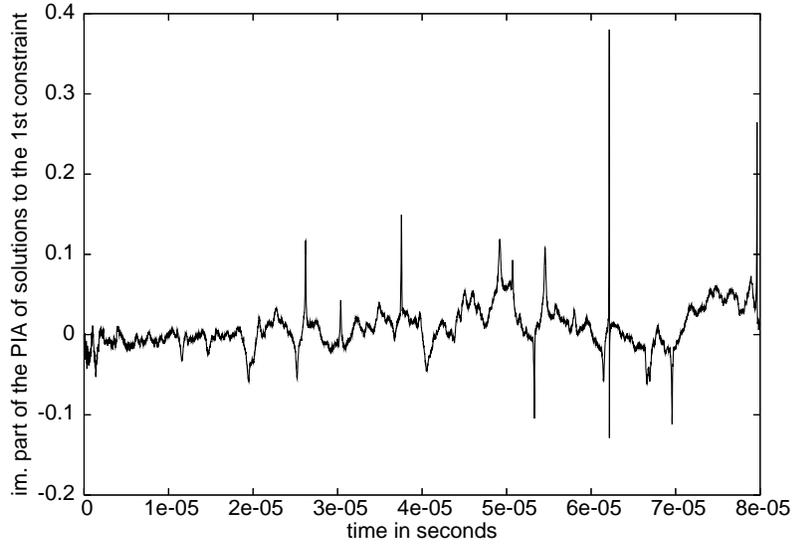}
\caption {The imaginary part of the path integral average (PIA) of
solutions to the first constraint equation \eqref{Ll2AAC} with zero
source rate ($J_H=J_{H_2}=0 s^{-1}$) for the hydrogen atoms under
interstellar conditions ($K=1.45 \times 10^5 s^{-1}$, $L_H=1.88
\times 10^{-3}s^{-1}$) over 1000 paths.}
\end{figure}

\begin{figure}[p]
\centering
\includegraphics[width=0.7\textwidth]{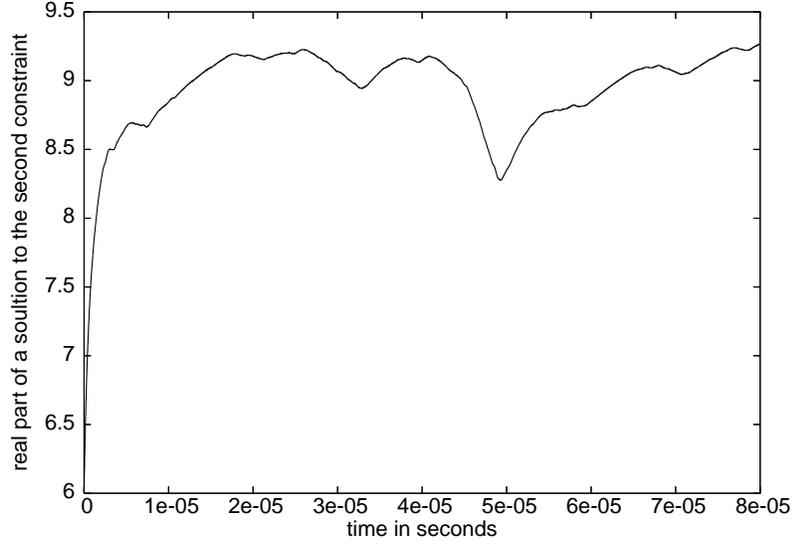}
\caption {The real part of a possible solution to the second
constraint equation \eqref{LlAAC} for vanishing source rate
($J_H=J_{H_2}=0 s^{-1}$), that is the constraint equation for the
diatomic hydrogen under interstellar conditions ($K=1.45 \times 10^5
s^{-1}$, $L_{H_2}=6.9 \times 10^{-8}s^{-1}$).}
\end{figure}

\begin{figure}[p]
\centering
\includegraphics[width=0.7\textwidth]{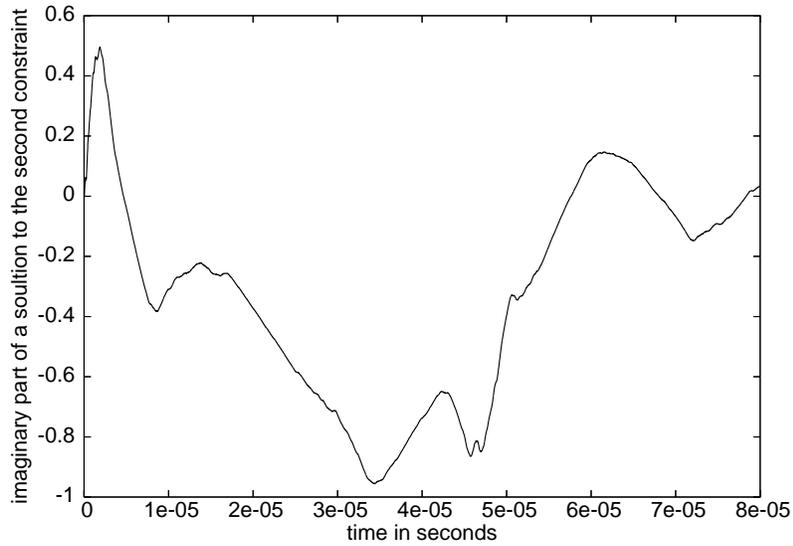}
\caption {The imaginary part of a possible solution for one specific
realisation of the stochastic noise to the second constraint
equation \eqref{LlAAC} with zero source rate ($J_H=J_{H_2}=0
s^{-1}$) under interstellar conditions ($K=1.45 \times 10^5 s^{-1}$,
$L_{H_2}=6.9 \times 10^{-8}s^{-1}$).}
\end{figure}

\begin{figure}[p]
\centering
\includegraphics[width=0.7\textwidth]{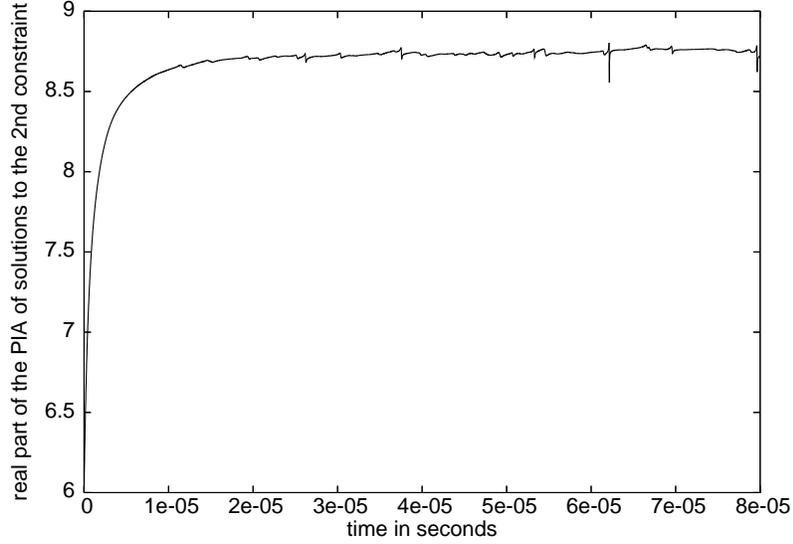}
\caption {The real part of the path integral average (PIA) of
solutions to the second constraint equation \eqref{LlAAC} for the
reaction product $H_2$ under interstellar conditions ($K=1.45 \times
10^5 s^{-1}$, $L_{H_2}=6.9 \times 10^{-8}s^{-1}$) over 1000 paths
with vanishing source rate ($J_H=J_{H_2}=0 s^{-1}$).}
\end{figure}
\begin{figure}[p]
\centering
\includegraphics[width=0.7\textwidth]{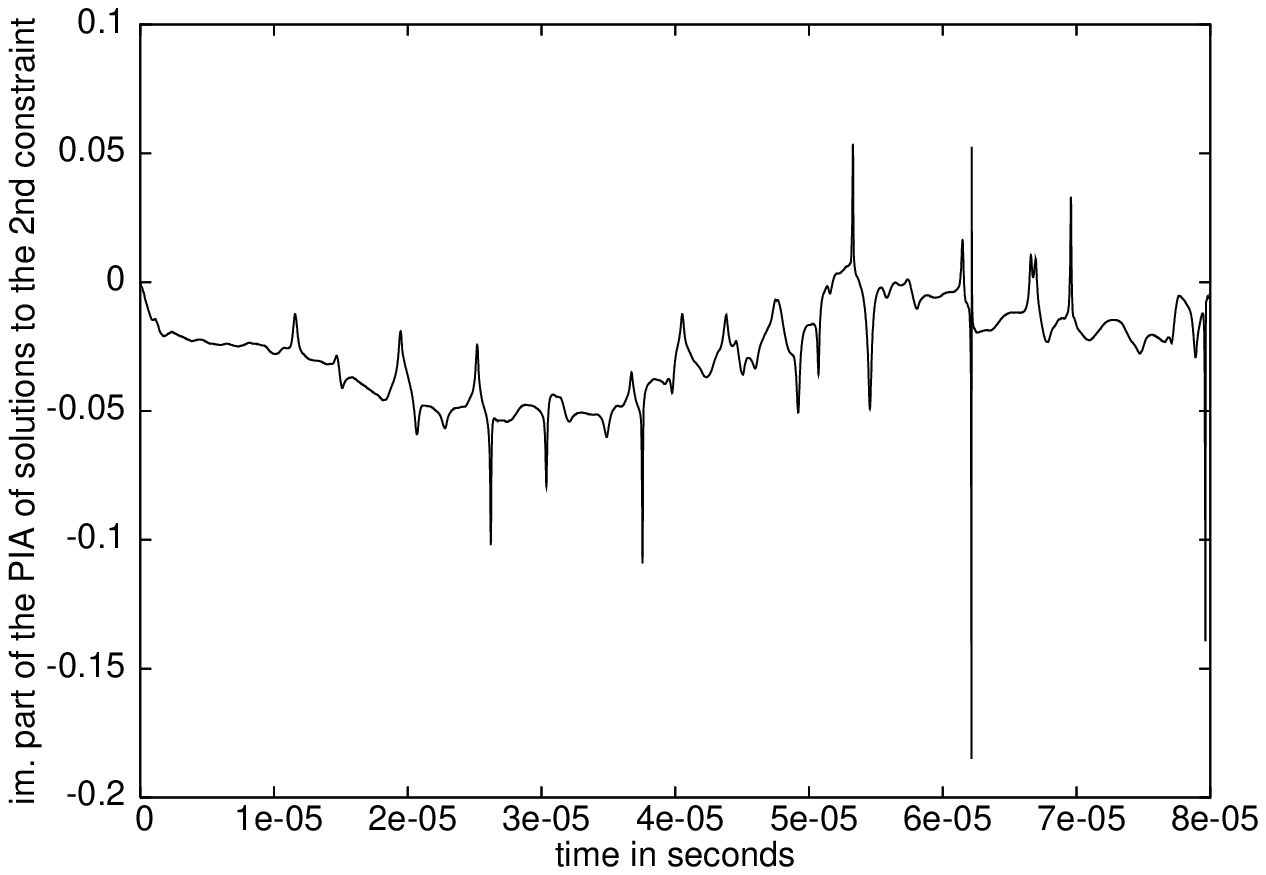}
\caption {The imaginary part of the path integral average (PIA) of
solutions to the second constraint equation \eqref{LlAAC} for the
diatomic hydrogen under interstellar conditions ($K=1.45 \times 10^5
s^{-1}$, $L_{H_2}=6.9 \times 10^{-8}s^{-1}$) over 1000 realisations
of the stochastic noise for zero source rate ($J_H=J_{H_2}=0
s^{-1}$).}
\end{figure}

\section{Case B: non-vanishing source rate}

As in the previous section, we generate solutions to the constraint
equations \eqref{Ll2AAC} and \eqref{LlAAC} in zero dimensions by the
numerical schemes discussed above but now with a non-zero source
rate. We compare the results to the solutions of the classical
evolution equations which read
\begin{equation}
\label{classAJ} n_A(t)=\frac{-\bar{\lambda}_A+\alpha
\text{tanh}(\frac{\alpha}{2} (t + \beta))}{4 \bar{\kappa}},
\end{equation}
\noindent where $\alpha:=\sqrt{8\bar{\kappa} \bar{j}_A +
\bar{\lambda}_A^2}$ and $\beta:=2/\alpha \
\text{arctanh}\big((4\bar{\kappa} n_A(0)+\bar{\lambda}_A)/\alpha
\big)$ and
\begin{equation}
 n_C(t)=e^{-\bar{\lambda}_C t}\big( \int_0^t
e^{\bar{\lambda}_C s}(\bar{j}_C + \bar{\kappa} n_A^2(s))ds +
n_C(0)\big).
\end{equation}
\noindent We now analyse the influence of the source rate
coefficient associated with the adsorption of reactants onto the
surface of the seed particle. For convenience, we will leave the
source rate for the reaction products at zero as it will not have
significant influence on the outcome of our discussion. The other
rate coefficients were chosen as before, $J_{A,C}$, $L_{A,C}$ and
$K_{A,C}$ denoting the rate coefficients per unit time. The plots in
Figures 9 to 12 were obtained for a source rate that is big in
comparison to the other rate coefficients, $J_H=10^8$ $s^{-1}$,
whereas in Figures 13 to 16 the source rate was chosen to be small,
$J_H=10^{-7}$ $s^{-1}$. Although one observes fluctuations both in
the real part of a single solution to the first constraint equation
(Figure 9) and in the real part of one path associated with the
second constraint equation (Figure 11), Figures 9 to 12 reproduce
deterministic behaviour. The real part of the path integral average
of the reactants (Figure 10), as well as the real part of the path
integral average of the reaction products (Figure 12) over 1000
possible realisations of the stochastic noise, coincide with the
results of the associated classical equations. This accordance
between classical and stochastic model is no longer valid for
Figures 13 to 16. As an example, we generated one single path for
each particle population, the reaction partners and the reaction
products, and plotted their real part in Figure 13 and Figure 15,
respectively. \\ \\
Let us now compare the average particle
populations of the hydrogen atoms and the diatomic hydrogen for
large source rate (Figure 10 and Figure 12) to the average particle
populations of the reactants and the reaction product for small
source rate (Figure 14 and Figure 16). In the deterministic case,
that is for large source rate, the chemical reaction does not die
out after a certain period of time
---see Figure 12--- in contrary to the observations made from Figure
16 where the real part of the path integral average of the diatomic
hydrogen over 1000 paths is plotted for a small source rate. As can
be seen from Figure 10 and Figure 14 respectively, for a source rate
of $J_H=10^8$ $s^{-1}$ the average particle population of the $H$
atoms reaches an asymptotic value of $22.28$ whereas for a source
rate of $J_H=10^{-7}$ $s^{-1}$ the average particle population of
the $H$ atoms is $1/2$ as was the case for vanishing source rate in
the latter section. Figure 12 and Figure 16 give the average
particle population for the $H_2$ atoms at time $t=8\times 10^{-5}$
$s^{-1}$, namely $3996.89$ for large source rate $J_H=10^8$ $s^{-1}$
and $8.73$ for small source rate $J_H=10^{-7}$ $s^{-1}$. To
determine the transition from a deterministic to a stochastic
behaviour we computed the average particle population of the
reaction partners and products for a source rate of the hydrogen
atoms in the range of $J_H \in [ 10^8 \ s^{-1},10^{-7} \ s^{-1} ]$
for each order of magnitude and generated Figure 17 to Figure 19.
Already when the reaction rate and the source rate are of the same
order of magnitude one can observe deviations from the classical
behaviour, in the sense that the equilibrium value of the average
particle population obtained by the stochastic methods is higher
than that predicted from the classical equation \eqref{classAJ}. One
also observes from Figure 17 that for a source rate between
$J_H=10^3$ $s^{-1}$ and $J_H=10^2$ $s^{-1}$ the chemical reaction
dies out because there is not one pair of reaction partners left in
order to initiate a chemical reaction.

\begin{figure}[p]
\centering
\includegraphics[width=0.7\textwidth]{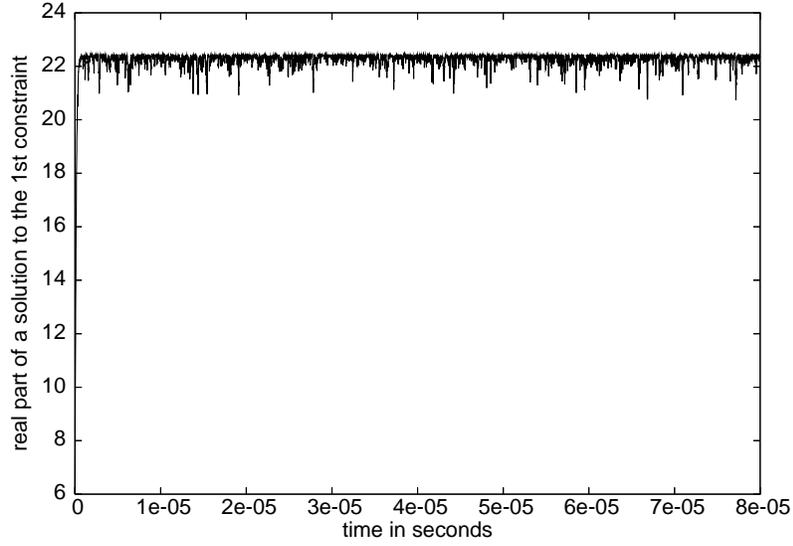}
\caption {Real part of one possible solution to the first constraint
equation \eqref{Ll2AAC} for a value of the source rate of $J_H=10^8$
$s^{-1}$ for the reactants and for $K=1.45 \times 10^5 s^{-1}$,
$L_H=1.88 \times 10^{-3}s^{-1}$.}
\end{figure}

\begin{figure}[p]
\centering
\includegraphics[width=0.7\textwidth]{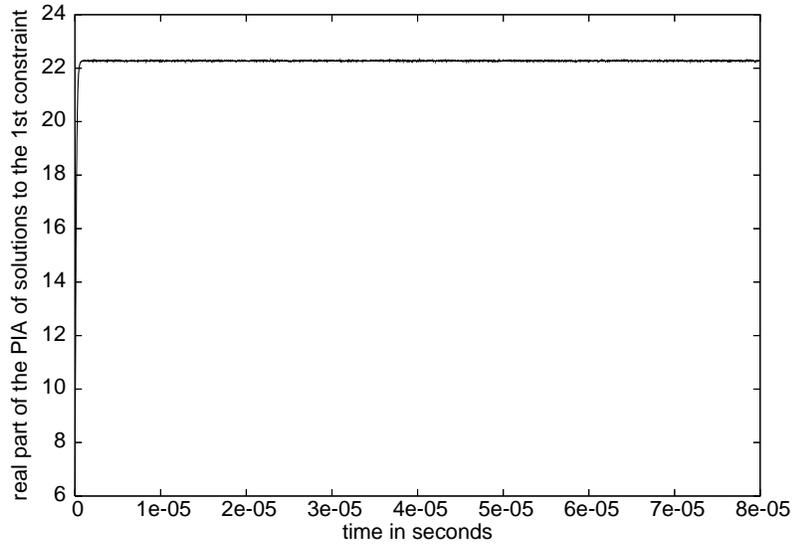}
\caption {Real part of the path integral average (PIA) of solutions
to the constraint equation \eqref{Ll2AAC} with $J_H=10^8$ $s^{-1}$
for the hydrogen atoms and $K=1.45 \times 10^5 s^{-1}$, $L_H=1.88
\times 10^{-3}s^{-1}$ averaged over 1000 paths.}
\end{figure}

\begin{figure}[p]
\centering
\includegraphics[width=0.7\textwidth]{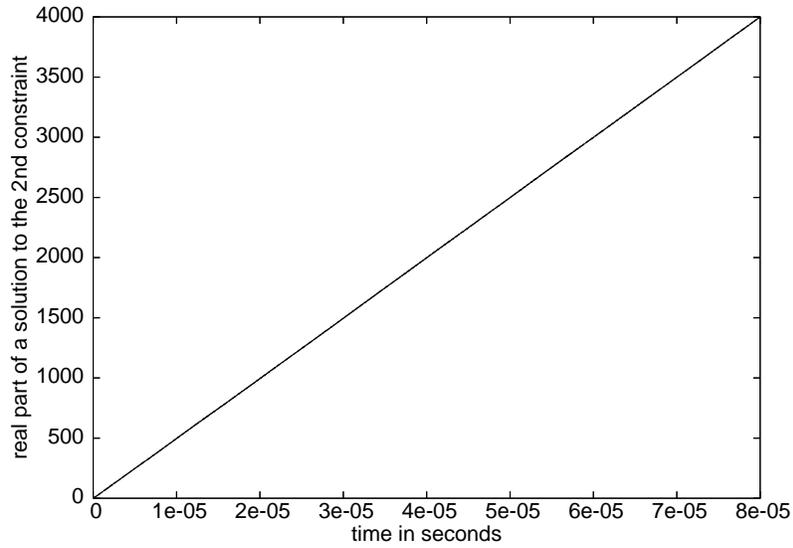}
\caption {Real part of one path for the constraint equation
\eqref{LlAAC} associated with the reaction product $H_2$ plotted for
the source rate $J_H=10^8$ $s^{-1}$ and $K=1.45 \times 10^5 s^{-1}$,
$L_{H_2}=6.9 \times 10^{-8}s^{-1}$, $J_{H_2}=0s^{-1}$.}
\end{figure}

\begin{figure}[p]
\centering
\includegraphics[width=0.7\textwidth]{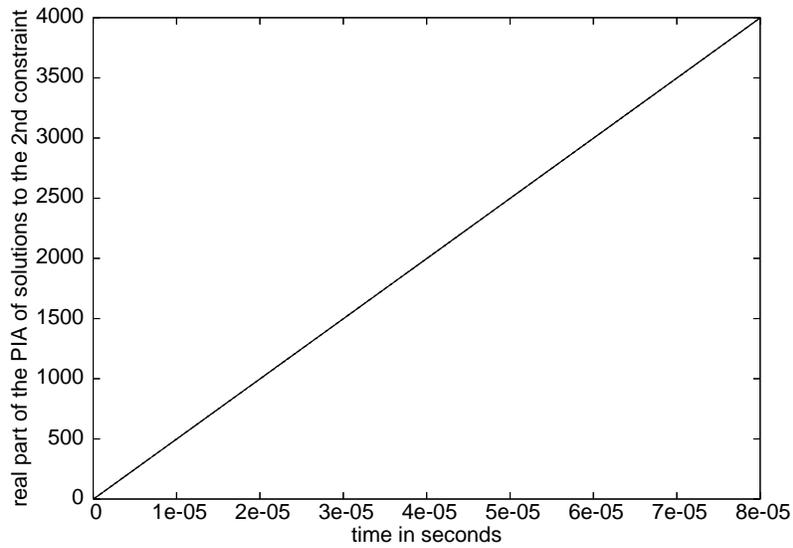}
\caption {Real part of the path integral average (PIA) for the
population of diatomic hydrogen for a large source rate compared to
the other rate coefficients, namely for $J_H=10^8$ $s^{-1}$, and for
$K=1.45 \times 10^5 s^{-1}$, $L_{H_2}=6.9 \times 10^{-8}s^{-1}$,
$J_{H_2}=0s^{-1}$.}
\end{figure}

\begin{figure}[p]
\centering
\includegraphics[width=0.7\textwidth]{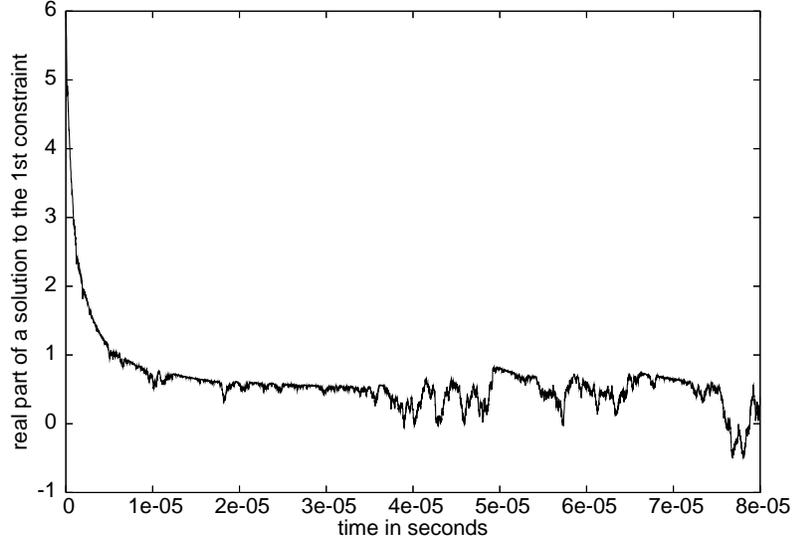}
\caption {Real part of a solution to the stochastic equation
\eqref{Ll2AAC} constraining the reaction partners ($K=1.45 \times
10^5 s^{-1}$, $L_H=1.88 \times 10^{-3}s^{-1}$) with a source rate
for the hydrogen atoms of value $J_H=10^{-7}$ $s^{-1}$.}
\end{figure}

\begin{figure}[p]
\centering
\includegraphics[width=0.7\textwidth]{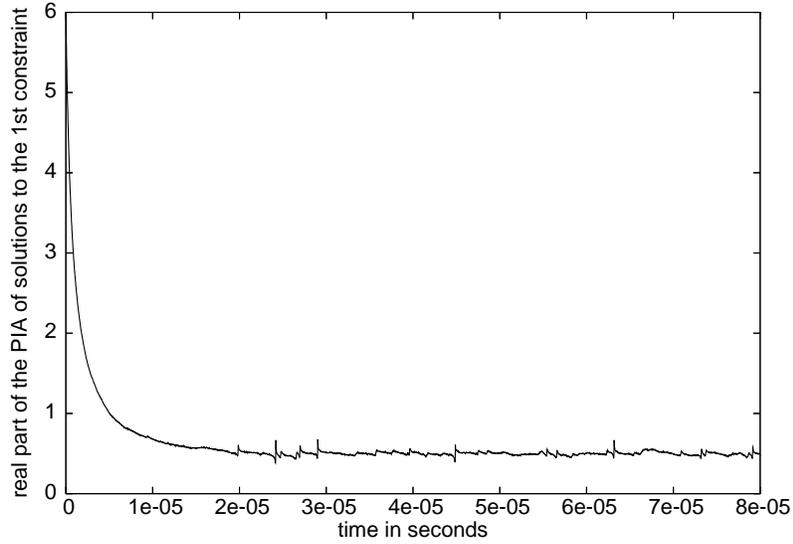}
\caption {Real part of the path integral average (PIA) of solutions
to the constraint equation \eqref{Ll2AAC} for the hydrogen atoms
over 1000 paths for a source rate of $J_H=10^{-7}$ $s^{-1}$ and for
$K=1.45 \times 10^5 s^{-1}$, $L_H=1.88 \times 10^{-3}s^{-1}$.}
\end{figure}

\begin{figure}[p]
\centering
\includegraphics[width=0.7\textwidth]{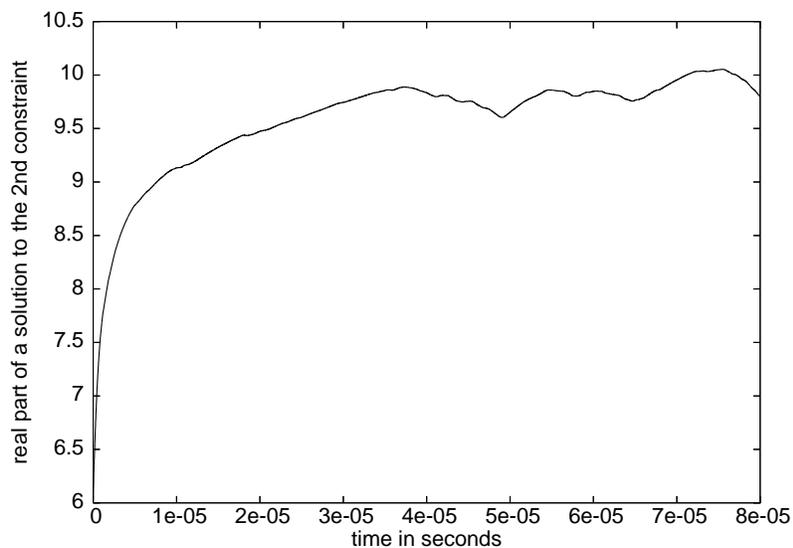}
\caption {Real part of one solution to the second constraint
equation \eqref{LlAAC} for a small source rate for the reactants
compared to the other rate coefficients: $J_H=10^{-7}$ $s^{-1}$, and
with $K=1.45 \times 10^5 s^{-1}$, $L_{H_2}=6.9 \times
10^{-8}s^{-1}$, $J_{H_2}=0s^{-1}$.}
\end{figure}

\begin{figure}[p]
\centering
\includegraphics[width=0.7\textwidth]{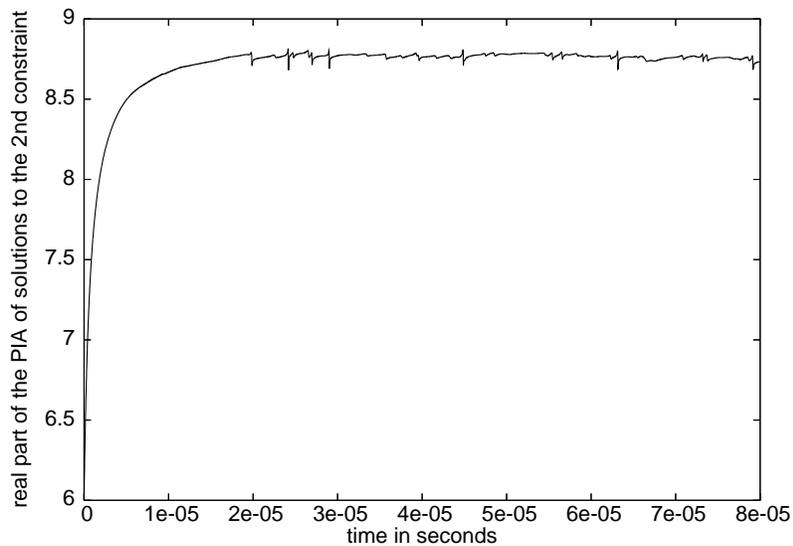}
\caption {Real part of the path integral average (PIA) of the
reaction products with $K=1.45 \times 10^5 s^{-1}$, $L_{H_2}=6.9
\times 10^{-8}s^{-1}$, $J_{H_2}=0s^{-1}$ over 1000 realisations of
the stochastic noise for a source rate of the hydrogen atoms of
$J_H=10^{-7}$ $s^{-1}$.}
\end{figure}

\begin{figure}[p]
\centering
\includegraphics[width=0.7\textwidth]{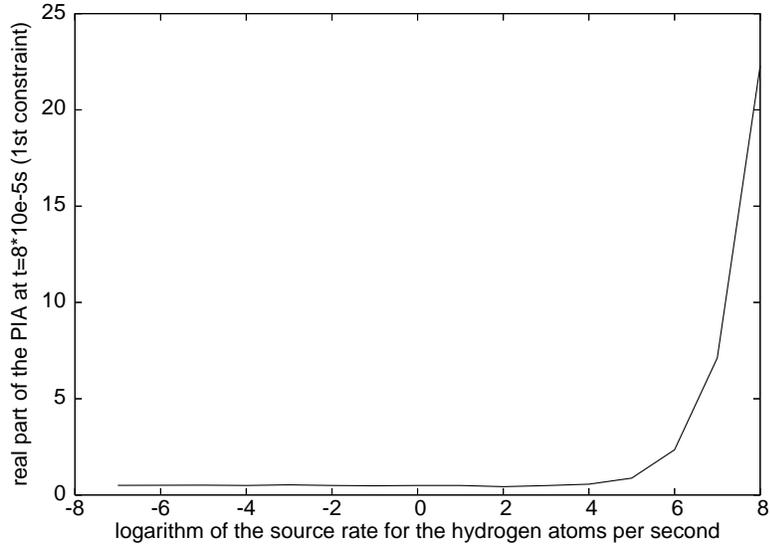}
\caption {Dependence of the real part of the path integral average
(PIA) of solutions to the first constraint equation \eqref{Ll2AAC}
taken over 1000 paths after the transient processes on the source
rate of the reactants for the average population of atomic hydrogen
($K=1.45 \times 10^5 s^{-1}$, $L_H=1.88 \times 10^{-3}s^{-1}$).}
\end{figure}

\begin{figure}[p]
\centering
\includegraphics[width=0.7\textwidth]{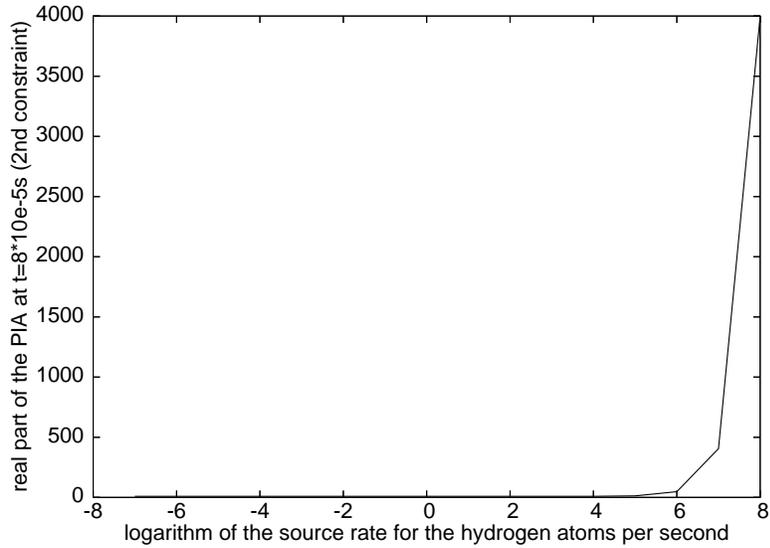}
\caption {Real part of the path integral average (PIA) of solutions
to the second constraint equation \eqref{LlAAC} ($K=1.45 \times 10^5
s^{-1}$, $L_{H_2}=6.9 \times 10^{-8}s^{-1}$, $J_{H_2}=0s^{-1}$) over
1000 paths for the average population of molecular hydrogen at
$t=8\times 10^{-5}$ $s$ versus the logarithm of the source rate of
the hydrogen atoms.}
\end{figure}

\begin{figure}[h]
\centering
\includegraphics[width=0.7\textwidth]{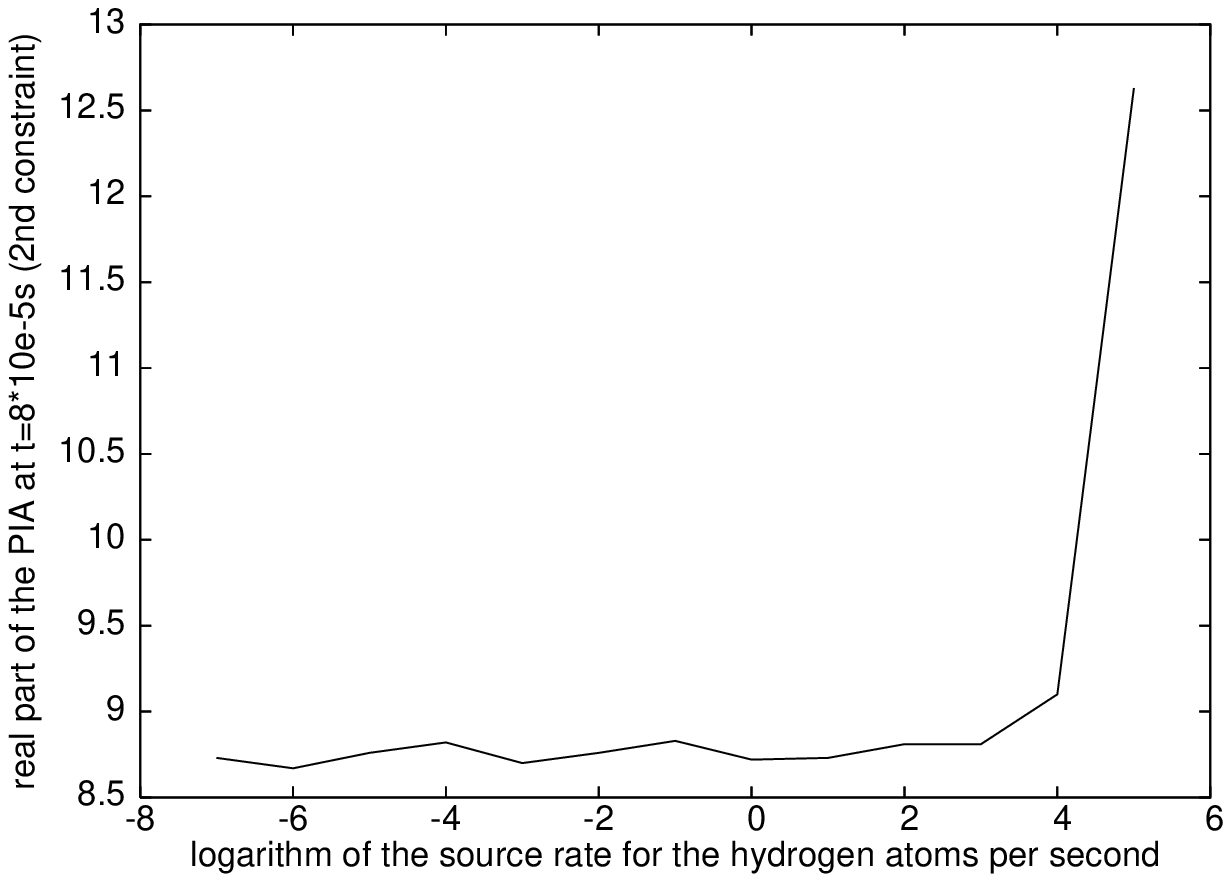}
\caption {Same curve as shown in Figure 18 but zoomed to resolve the
plot close to the horizontal axis.}
\end{figure}

\section{Conclusions}

We have argued that the classical evolution equations for the mean
particle population of a chemical species involved in a
heterogeneous chemical reaction do not give the right results for
small systems. Instead, we developed a stochastic model that
includes statistical fluctuations and showed in our numerical
investigations that those fluctuations can not be ignored for low
rates of particle adsorption onto the surface of a grain particle.
Although one starts from an apparently classical system the
introduction of a quantum field theoretical formalism for its
description seems to force us to adopt a "quantum mechanical-like"
interpretation of the results.\\ \\
It is possible to extend this work to other chemical reactions, for
example of the type $A+B \longrightarrow C$. In a next step, we
shall consider a network of chemical reactions in which several
reactions compete against each other. One would expect that it will
take considerably longer to reach an asymptotic value for the
average particle population of a certain species.


\section{Acknowledgements}

This work was supported by the Leverhulme Trust under grant
F/07134/BV and partly supported by the C N Davies Award of the
Aerosol Society.


\end{document}